\documentclass[amsmath,amssymb,prl,hyperlink,twocolumn]{revtex4}

	\usepackage{graphicx}
	\usepackage{soul}
	\usepackage[colorlinks=true,citecolor=blue,linkcolor=magenta]{hyperref}
	\usepackage[usenames]{color}
	\usepackage{amsfonts}
	\usepackage{color}
	\usepackage{booktabs}
	\usepackage{multirow}
	\usepackage{float}
    \usepackage{times}
    \usepackage[english]{babel}

\begin{document}

\title{{Emergence of Classical Objectivity of Quantum Darwinism in a Photonic Quantum Simulator}}

\author{Ming-Cheng Chen$^{1,2}$, Han-Sen Zhong$^{1,2}$, Yuan Li$^{1,2}$, Dian Wu$^{1,2}$, Xi-Lin Wang$^{1,2}$, Li Li$^{1,2}$, Nai-Le Liu$^{1,2}$}

\author{Chao-Yang Lu$^{1,2}$}  \email{cylu@ustc.edu.cn}
\author{Jian-Wei Pan$^{1,2}$ \vspace{0.2cm} }

\affiliation{$^1$ Hefei National Laboratory for Physical Sciences at Microscale and Department of Modern Physics, University of Science and Technology of China, Hefei, Anhui 230026, China}
\affiliation{$^2$ CAS Centre for Excellence and Synergetic Innovation Centre in Quantum Information and Quantum Physics, University of Science and Technology of China, Hefei, Anhui 230026, China.}
\affiliation{}
\date{\today}

\begin{abstract}
	Quantum-to-classical transition is a fundamental open question in physics frontier. Quantum decoherence theory points out that the inevitable interaction with environment is a sink carrying away quantum coherence, which is responsible for the suppression of quantum superposition in open quantum system. Recently, quantum Darwinism theory further extends the role of environment, serving as communication channel, to explain the classical objectivity emerging in quantum measurement process. Here, we used a six-photon quantum simulator to investigate  {classical and quantum information proliferation} in quantum Darwinism process.  {In the simulation, many environmental photons are scattered from an observed quantum system and they are collected and used to infer the system's state}. We observed redundancy of system's classical information and suppression of quantum correlation in the fragments of environmental photons. Our results experimentally  {show that the classical objectivity of quantum system can be established through quantum Darwinism mechanism.}

	\end{abstract}
	\maketitle


~\\

\begin{figure}[t]
	\centering
	\includegraphics[width=0.4\textwidth]{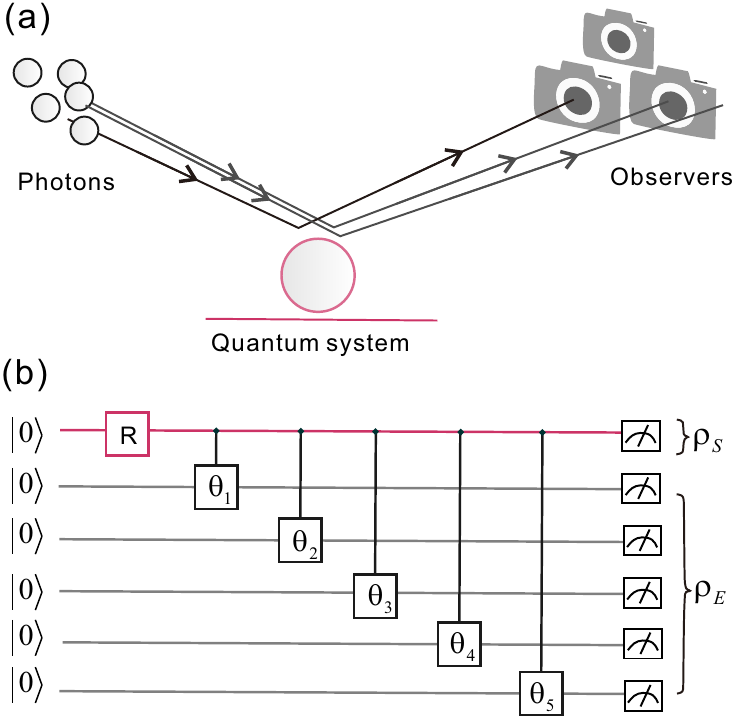}
	\caption{Quantum Darwinism process. (a) Multiple observers use independent fragments of  the scattered environment photons to reveal the state of observed quantum system. They can determine the pointer states of the observed quantum system without perturbing it and thus agree on the observed outcome. As a result, the quantum system becomes classical and objective in this process. (b) A quantum simulator to simulate the system-environment interaction and produce quantum Darwinism states. In the simulator,  {the fist qubit is quantum system and the environment particles (other qubits) interact with the system of arbitrary interaction strengths $\{\theta _i\}$ in parallel and have no interaction between themselves. Measurements are performed on these particles to infer the information of quantum system.} }
	\label{fig1}
\end{figure}

\begin{figure}[tb]
	\centering
	\includegraphics[width=0.4\textwidth]{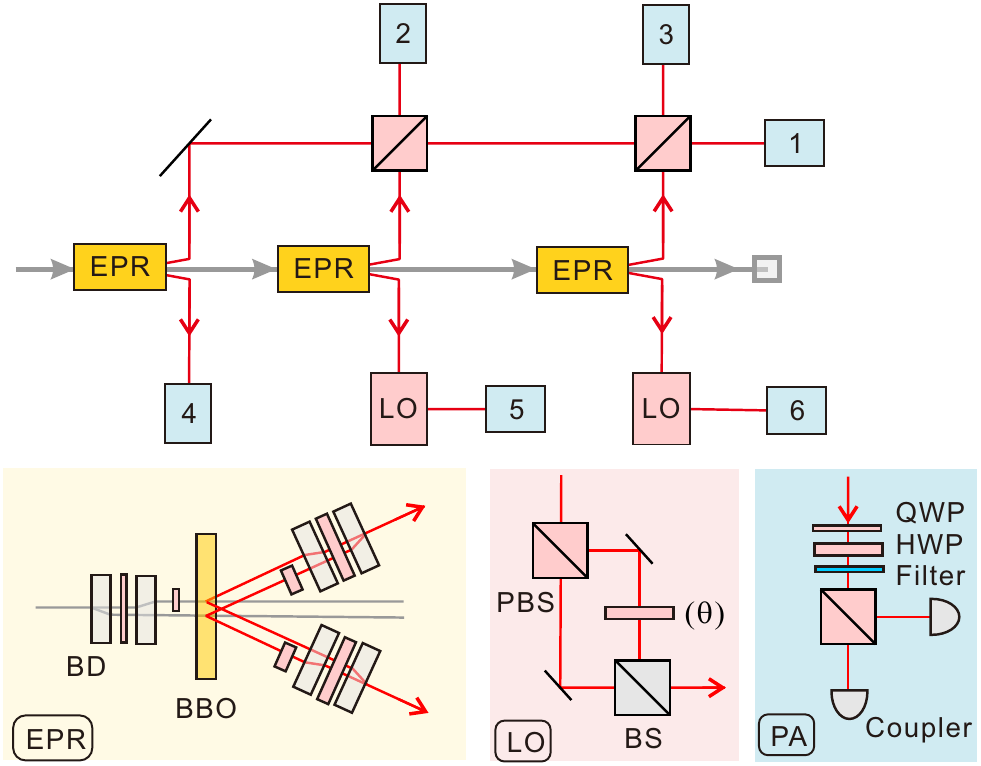}
	\caption{Experimental setup. A six-photon interferometer is used to produce the system-environment composite quantum Darwinism states. Photon 1 is the central quantum system and photons $2 \sim 6$ are the environment. Infrared laser pulses (775 nm wavelength, 120 fs pulse duration, 80 MHz repetition rate) pass through three beta-barium  borate (BBO) nonlinear crystals sequentially to produce three pairs of Einstein-Podolsky-Rosen(EPR) polarization-entangled photons. The two components of EPR state are independently produced and coherently combined by beam displacers (BDs). Three single photons, one from each pair (photons 1, 2 and 3), interfere on two polarization beam splitters (PBSs) to generate an entangled six photons in Greenberger-Horne-Zeilinger (GHZ) state. Two single photons (5 and 6) pass through polarization-dependent Mach-Zehnder (MZ) interferometers to realize local non-unitary operations (LOs). In the MZ interferometers, two polarization components of input photons are separated by polarization beam splitters and recombined on balanced beam splitters (BSs). The internal half-wave plates (HWP) are set at angle ${\theta _5/4}$ and  ${\theta _6/4}$, respectively. All the photons are  sent to polarization analysis (PA) setups, each consisting of a quarter-wave plate (QWP), a HWP, a spectrum filter, and a PBS. The photons are finally detected by fiber-coupled single-photon detectors and six-fold coincidence counting are registered.  {Note that our experimental setup is not a faithful simulator of the scattering process. Instead, we mainly want to simulate the process where scattered photons are used to infer the state of quantum system.}}
	\label{fig2}
\end{figure}

\begin{figure*}[t]
	\centering
	\includegraphics[width=1\textwidth]{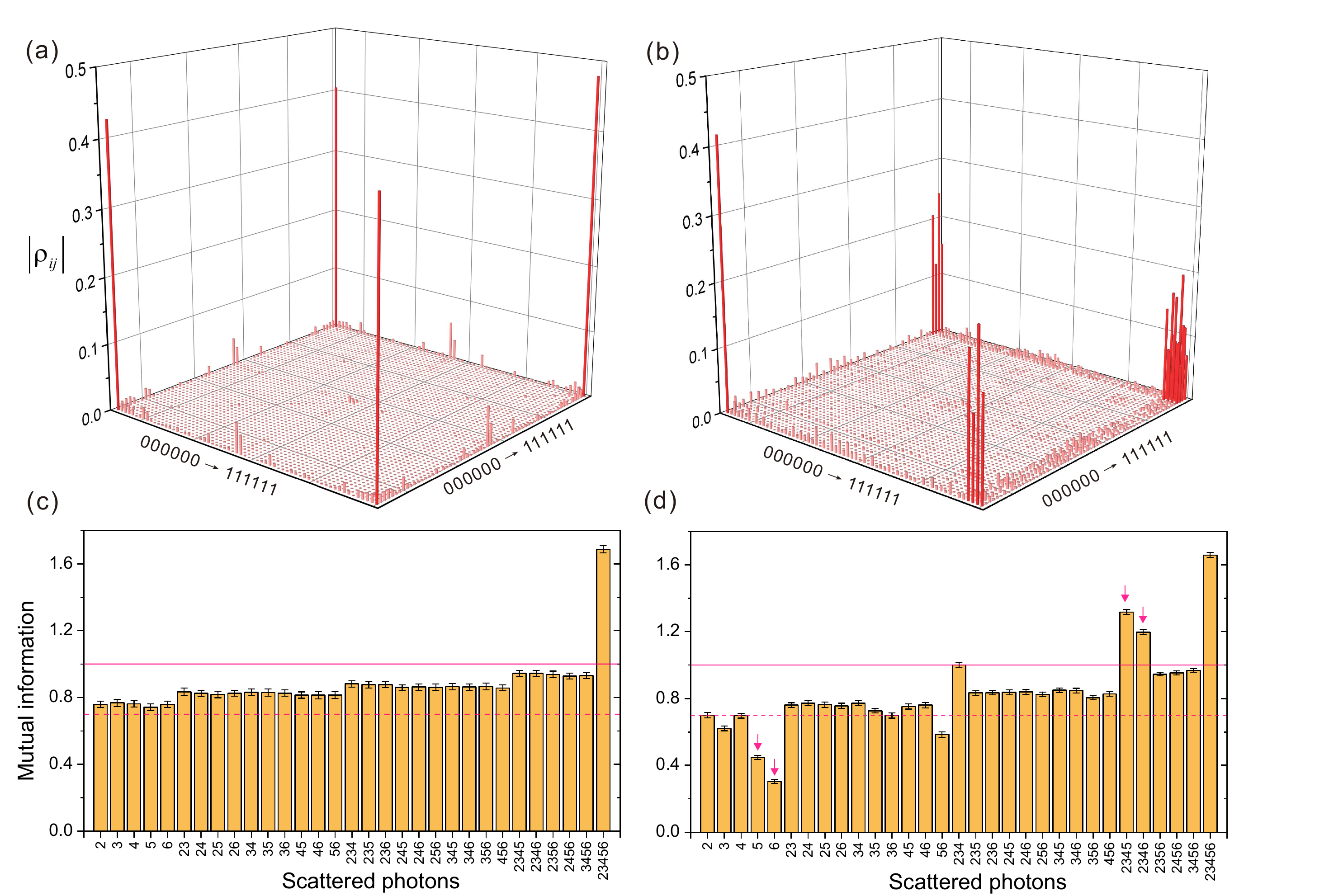}
	\caption{Experimental results. (a)(b) The absolute value of restructured density matrix elements of Darwinism states for parameter ${\vec \theta _A}$ and ${\vec \theta _B}$, respectively. The states are estimated by quantum state tomography with $729$ measure settings. (c)(d) The mutual information between the system and the $31$ different pieces of environment fragments for parameter ${\vec \theta _A}$ and ${\vec \theta _B}$, respectively.  The red line indicates the classical entropy of system and the dotted line marks $70\%$ of the classical entropy to guide the eye. The  short arrows label four special environment fragments for further analysis in the main text.
   }
	\label{fig3}
\end{figure*}

\begin{figure}[t]
	\centering
	\includegraphics[width=0.45\textwidth]{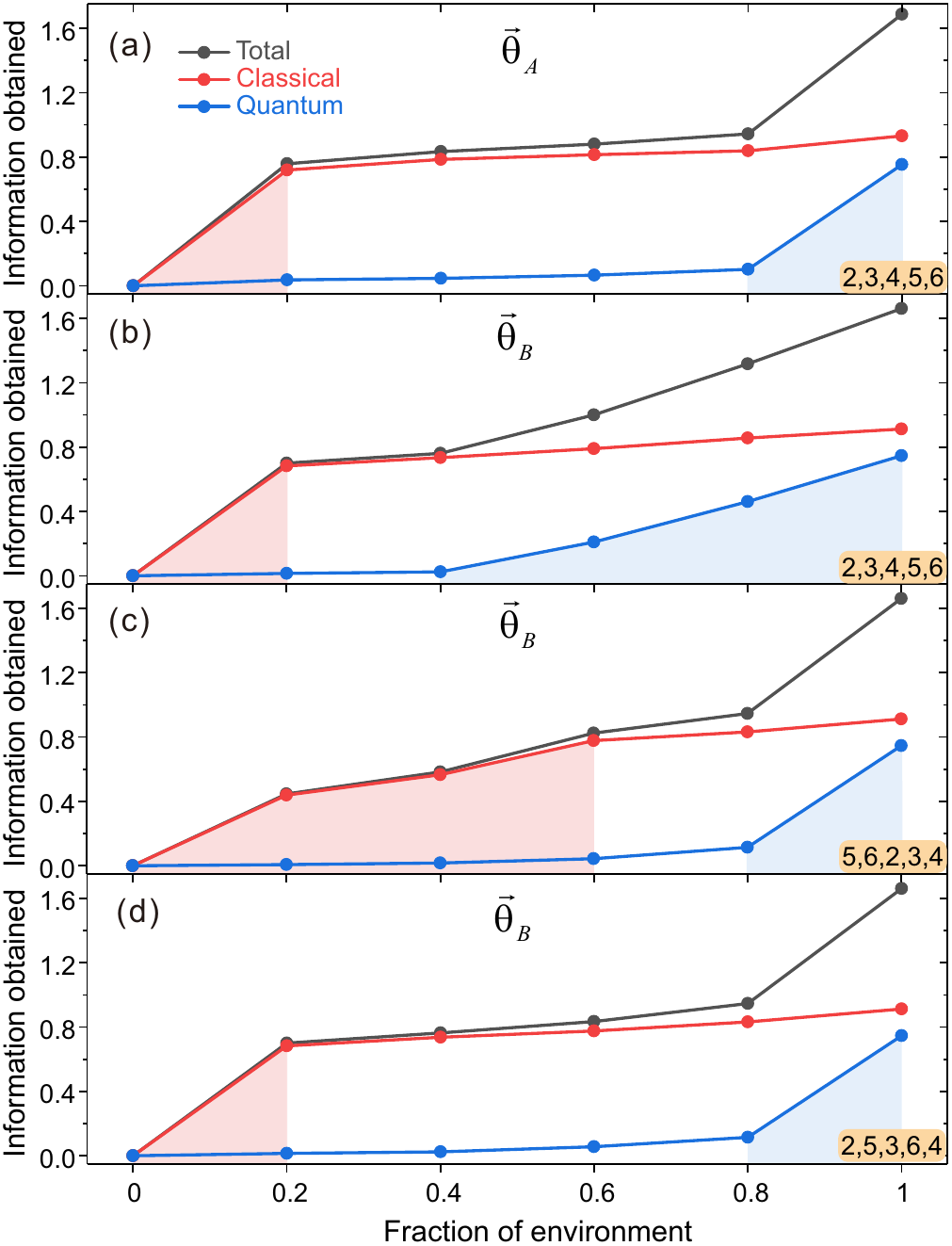}
	\caption{Accessible information of local observers. (a) Quantum Darwinism process ${\vec \theta _A}$. The total correlation, classical correlation (Holevo bound) and pure quantum correlation (quantum discord) between the system and the environment are shown. The fraction of environment increase in the order of photons 2,3,4,5 and 6. (b)(c)(d) Quantum Darwinism process ${\vec \theta _B}$. The orders of environment photons are 23456, 56234 and 25364, respectively. One standard deviation is  {smaller} than the size of data marker. Redundancy plateau of classical information correlation in environment fragments is observed meanwhile pure quantum correlation is suppressed due to the incompleteness of environment. Local observers use only small fraction of environment to infer the system, so only classical information is accessible.
 }
	\label{fig4}
\end{figure}

\section{Introduction}

Quantum mechanics is a spectacularly successful predictive theory, but there is still an unresolved problem about its interpretation in quantum measurement problem \cite{decoherence2005rmp}.  {The orthodox Copenhagen interpretation separates the world into quantum domain and classical domain, which is bridged by observation-induced collapse 
\cite{collapse2013rmp}.
 How the wave function collapses and classical objectivity emerges from a quantum substrate?}
  A detailed mechanism of this quantum-to-classical transition is of fundamentally importance for developing a unified view of our physical world. 

Quantum decoherence theory identifies that the uncontrolled interactions  {with} the environment can destroy the coherence of  {a} quantum system into a mixed state.  {In the theory, the environment is traced out and thus the system's classical behavior is explained in the level of ensemble average \cite{decoherence1985,decoherence2003rmp}. How the quantum system's classical objectivity arises in a single measurement event is still unresolved. Classical objectivity is a property that many observers can independently observe and establish a
consensus view of the state of a quantum system without perturbing it \cite{darwinism2009nphy}. In a general observation process,} observers don't directly touch and interact with the quantum system. They perceive the system by collecting information from its surrounding environment. 

Recently, quantum Darwinism explains the emergence of classical objectivity of a single quantum system through  {classical} information broadcasting and proliferating in its environment \cite{darwinism2009nphy,darwinism2018}. The key idea is that the environment acts as communication channel and only  {classical} information about the system can reach observers. The environment selects system's  {classical} information to broadcast and proliferate, and observers use the redundant  {classical} information in local environment fragments to perceive the state of system. In this process, many observers can independently and simultaneously query separate fragments of the environment and reach a consensus about the system's  {classical} state.  {Specifically, the quantum Darwinism theory singles out a branch structure of system-environment(observers) composite quantum states \cite{Brownian2016,darwinism2006pra,darwinism2015} from measurement-like interaction to explain the appearance of classical pointer states. The classical pointer states are the eigenstates of the measurement observable.} 
 
In this work, we report a test of quantum Darwinism principle on a photonic quantum simulator \cite{simulator1996,simulatorPhoton} in view of information theory. We measured the information correlations between system and environment {, where the system is a single photon and the environment is another five photons}.  {Quantum mutual information, Holevo bound, and quantum discord  \cite{discord2003prl,discord2011prl,discord2012rmp,discord2013srep} are used }to account for the total correlation, classical correlation, and pure quantum correlation, respectively. We used these correlations to investigate information broadcasting and proliferating.

\section{Theory}
The basic process of quantum Darwinism is shown in Fig.\ \ref{fig1}(a). A central quantum system  {(single photon)} is monitored by particles  {(photons)} in the environment \cite{envEveryday2010prl,envNoisy2014prl}. The particles are scattered from the system and caught by observers. These environment particles serve as individual memory cells which are imprinted of system's pointer-state information. When there are random interactions among the environment's particles, the stored information will be inevitably scrambled out. Hence, only non-interaction environment is good memory for redundant records of system's state. 
 
We design a quantum Darwinism simulator shown in Fig.\ \ref{fig1}(b) to simulate non-interaction environment, such as daily photonic environment. In the simulator, a central qubit interacts with the environment qubits through two-qubit controlled-rotation gates $U(\theta ) = \left| 0 \right\rangle \left\langle 0 \right| \otimes I + \left| 1 \right\rangle \left\langle 1 \right| \otimes {R_y}(\theta )$ with random angles to mimic the random scattering process, where ${R_y}(\theta )$ rotates a qubit by angle $\theta $ along the $y$ axis of Bloch sphere. When the system qubit is initialized in superposition state $\alpha {\left| 0 \right\rangle _S} + \beta {\left| 1 \right\rangle _S}$, the simulator will produce  {Darwinist} states with branch structure  
\begin{align} 
	\alpha {\left| 0 \right\rangle _S} \otimes _{i = 1}^N{\left| 0 \right\rangle _i} + \beta {\left| 1 \right\rangle _S} \otimes _{i = 1}^N(\cos \frac{{{\theta _i}}}{2}{\left| 0 \right\rangle _i} + \sin \frac{{{\theta _i}}}{2}{\left| 1 \right\rangle _i}) \label{state}
,\end{align} where ${\left| \alpha  \right|^2} + {\left| \beta  \right|^2} = 1$ and $N$ is the number of environment qubits. 

The interaction with environment selects preferred pointer states of the observed system, which are the states left unchanged under the interactions and thus multiple records of the state can be faithfully copied into environment. For interactions generated from a Hamiltonian form of ${H_i} = {g_i}A \otimes {B_i}$, the eigenstates of monitored observable $A$ are the pointer states, where $A$ and ${B_i}$ are two observables on system and environment particle $i$, respectively  {\cite{darwinism2009nphy}}. In our simulator setting, $A = ({\sigma _I} - {\sigma _Z})/2$, ${B_i} = {\sigma _X}$ and ${g_i}\Delta t = {\theta _i}/2$, therefore, the pointer states from interaction $U({\theta _i}) = {e^{ - i{H_i}\Delta t}}$ are $\left| 0 \right\rangle$ and $\left| 1 \right\rangle $, respectively.

The pointer states can be quantified by the disappearance  of quantum coherence 
\begin{align}
	C({\rho _S}) = {H_{cl}}({\rho _S}) - H({\rho _S}) \label{coherence}
\end{align}
where ${\rho _S}$ is the reduced density matrix of system, $H({\rho _S}) =  - tr({\rho _S}{\log _2}{\rho _S})$ is quantum von-Neumann entropy and ${H_{cl}}({\rho _S}) =  - tr({{\bf{p}}_s}{\log _2}{{\bf{p}}_s})$ is classical Shannon entropy, ${{\bf{p}}_s}$ is the diagonal elements of density matrix ${\rho _S}$  in pointer-state bases \cite{coherenceQuantify,coherenceRMP}. The reality of pointer states will emerge when the quantum system is completely decohered by the environment. In this case, the classical entropy will equal to the quantum entropy. The efficiency of decoherence depends on the initial states of environment \cite{envNonideal2010pra,envNoisy2014prl}. Impure or misaligned (close to the eigenstate of observable ${B_i}$) environment will reduce the decoherence efficiency. In our simulation, $\left| 0 \right\rangle $ states are used as initial environment states with optimal efficiency. 

In quantum Darwinism process, the information about the quantum system is broadcast to the environment. Local observers can only access small fragments of the whole environment.The quantum mutual information \begin{align}
	I(S\mathpunct{:}{E_i}) = H({\rho _S}) + H({\rho _{{E_i}}}) - H({\rho _{S{E_i}}}) \label{mutual}
\end{align}
 can be used to quantify how much information an observer ${E_i}$ (accessing the environment fragment ${E_i}$) knows about the quantum system \cite{darwinism2009nphy}. When $I(S\mathpunct{:}{E_i}) \approx H({\rho _S})$ for all observers $\{ {E_i}\} $, the system's state can be determined by all the observers and thus the quantum system becomes objective.

\section{Experiment}
We use a photonic simulator \cite{simulator1996,simulatorPhoton} (shown in Fig.\ \ref{fig2}) to produce quantum Darwinism states consisted of a central system qubit (photon 1) and five environment qubits (photons $2\sim6$ ) . The quantum state of system is observed at the superposition state $(\left| 0 \right\rangle  + \left| 1 \right\rangle )/\sqrt 2 $. Two sets of rotation-angle parameter, ${\vec \theta _A} = ({180^ \circ },{180^ \circ },{180^ \circ },{180^ \circ },{180^ \circ })$ and ${\vec \theta _B} = ({180^ \circ },{180^ \circ },{180^ \circ },{72^ \circ },{100^ \circ })$, are used in the experiments from the following considerations.  {Note that here the phases in ${\vec \theta _B}$ are chosen merely to represent nonorthogonal case to simulate the small environment fragment without specific optimization.}

A real environment fragment can contain many elementary subsystems. If the fragment is large, its quantum states can be simplified and expressed as orthogonal logical states $\left| {{0_L}} \right\rangle  =  \otimes _{i = 1}^n{\left| 0 \right\rangle _i}$ and $\left| {{1_L}} \right\rangle  =  \otimes _{i = 1}^n(\cos \frac{{{\theta _i}}}{2}{\left| 0 \right\rangle _i} + \sin \frac{{{\theta _i}}}{2}{\left| 1 \right\rangle _i})$, due to $\left\langle {{{0_L}}}
\mathrel{\left | {\vphantom {{{0_L}} {{1_L}}}}
	\right. \kern-\nulldelimiterspace}
{{{1_L}}} \right\rangle  = \prod _{i = 1}^n\cos \frac{{{\theta _i}}}{2} \to 0$ for sufficiently large number $n$ of subsystems in a fragment. Furthermore, if the fragment is small, its quantum states can  be expressed as nonorthogonal logical states $\left| {{0_L}} \right\rangle$ and $\cos \frac{\theta }{2}\left| {{0_L}} \right\rangle  + \sin \frac{\theta }{2}\left| {{1_L}} \right\rangle$. In our simulation, the photonic qubits represent the quantum state of environment fragments logically and an observer can access one or more fragments to infer the system's state. The parameter ${\vec \theta _A}$ is used to simulate 5 large environment fragments and parameter ${\vec \theta _B}$ is used to simulate 3 large environment fragments and 2 small environment fragments.

The experimental setup is shown in Fig.\ \ref{fig2}. The qubits are encoded in the horizontal ($H$) and vertical ($V$) polarization of single photons, which are produced by spontaneous parametric down-conversion (SPDC) process \cite{spdc,multiphotonRMP}. The Darwinism states in equation (\ref{state}) are synthesized in three steps. 

	(1) Preparation of three pairs of polarization-entangled photons in Einstein-Podolsky-Rosen (EPR) state $(\left| {HH} \right\rangle  + \left| {VV} \right\rangle )/\sqrt 2 $ \cite{epr} from SPDC process. With a laser pump power of 0.9 W , the generation probability of two twin photons  is 0.025 and the fidelity of EPR state is above $99\%$.

	(2) Three EPR pairs are combined on two polarization beam splitters, postselecting the entangled subspace of $\left| {HH} \right\rangle \left\langle {HH} \right| + \left| {VV} \right\rangle \left\langle {VV} \right|$, to produce a six-photon Greenberger-Horne-Zeilinger (GHZ) state $(\left| {HHHHHH} \right\rangle  + \left| {VVVVVV} \right\rangle )/\sqrt 2 $ \cite{ghz}. The success probability is 0.25.

	(3) Two photons from the GHZ state are further operated by polarization-dependent Mach-Zehnder (MZ) interferometer. In the interferometer, the H and V components are separated by polarization beam splitters and then recombined on balanced beam splitters to implement non-unitary process $\left| H \right\rangle \left\langle H \right| + (\cos \frac{\theta }{2}\left| H \right\rangle  + \sin \frac{\theta }{2}\left| V \right\rangle )\left\langle V \right|$ with a success probability 0.5.

The experiment runs with repetition rate 80 MHz and the single photons are measured with collecting and detecting efficiency of 0.65. Thus, we obtain six-photon coincidence counting rates about $5$ counts/second. The quantum states are measured through quantum state tomography. There are 729 measurement settings and about $700$ counts in each setting. Fig.\ \ref{fig3}(a) and (b) show the measured density matrices. For setting ${\vec \theta _A}$, the quantum state fidelity is $0.859 \pm 0.002$ and the purity is $0.777 \pm 0.004$.  For setting ${\vec \theta _B}$, the quantum state fidelity is $0.703 \pm 0.004$ and the purity is $0.692 \pm 0.006$. The standard deviation is estimated from Monte Carlo method with 100 trials.

The quantum coherence $C({\rho _S})$ of system qubit in equation (\ref{coherence}) are $0.001(4)$ and $0.020(6)$ in process ${\vec \theta _A}$ and ${\vec \theta _B}$, respectively, which indicates that the environment has fully decohered the system. The quantum mutual information $I(S\mathpunct{:}{E_i})$, the equation (\ref{mutual}), between the system and 31 different combinations of environment fragments  are shown in the Fig.\ \ref{fig3}(c) and (d), respectively. The results have two significant features. The mutual information quickly approaches the system's entropy $H({\rho _S})$ (exceeding $70\%$, mainly limited by the purity of prepared entangled quantum states) when accessing small environment fragments, and the mutual information is saturated when increasing the size of environment fragments.

The arrows in Fig.\ \ref{fig3}(d) show that the environment fragments  $5$ and $6$ are two low-fidelity records of the system's states, which is expected from the corresponding nonorthogonal rotating angles in process ${\vec \theta _B}$. On the other hand, environment fragments $2346$ and $2345$  have mutual information exceeding the system's entropy. This indicates that the mutual information contains more information than system's information alone. We further analyze the information compositions by dividing it into the locally-accessible classical information and the extra information from pure quantum correlations. We show the classical correlation and quantum correlation between system and environment fragments in Fig.\ \ref{fig4}. 

The first one, Hovelo bound $\chi (S\mathpunct{:}{E_i})$, measures the capacity of environment acting as communication channel to deliver system's classical information.  The Hovelo bound
\begin{align}
	\chi (S\mathpunct{:}{E_i}) = \mathop {\max }\limits_{\{ {M_s}\} } \{ H(\sum\limits_s {{p_s}{\rho _{{E_i}|s}}} ) - \sum\limits_s {{p_s}H({\rho _{{E_i}|s}})} \} 
\end{align}
is maximum mutual information of the classical-quantum state between system and environment fragments with optimal measurement $\{ {M_s}\}$ on the system \cite{discord2012rmp}, where $\rho _{{E_i}|s}$ is quantum state of environment ${E_i}$ conditioned on a measured result $s$ on the system with probability $p_s$ . The second one, quantum discord 
\begin{align} 
D(S\mathpunct{:}{E_i}) = I(S\mathpunct{:}{E_i}) - \chi (S\mathpunct{:}{E_i})
\end{align}
 measures the loss of information due to the observers can only locally access the environment, which quantifies the pure quantum correlation between environment and system \cite{discord2012rmp}.

In Fig.\ \ref{fig4}, the classical correlations and quantum correlations display very different features. The classical correlations have initial rise and saturate at classical plateau, which indicate the environment has recorded redundant copies of system's classical information for independent observers. In sharp contrast, the quantum correlations raise when the nearly whole environment is accessed, manifesting quantum correlations cannot be shared between the observers \cite{noShared}. The Fig.\ \ref{fig4} (b) and (c) also demonstrate the effects of low-fidelity environment fragments (photons 5 and 6 in process ${\vec \theta _B}$ ), which will lead to early raise of quantum correlation (Fig.\ \ref{fig4} (b)) or delay the raise of classical correlation (Fig.\ \ref{fig4} (c)).

\section{Discussion}
Our results exhibit that environment not only decoheres quantum system  {but} also selectively delivers the system's information to observers. The environment channel is high-efficient for classical information and inefficient for quantum information. Only the classical information of quantum system's decohered pointer states survives the environment-selected broadcasting and proliferates throughout the environment. Consequently,  {these results show that Quantum Darwinism theory predictions are compatible with the observation that classical objectivity originates from Darwinism-like broadcast structure of quantum substrate and \textit{quantum objectivity} is prohibited by quantum mechanics due to quantum no-broadcast phenomenon \cite{noBroadcast,noShared}.}

 {In the experiment, the observation of quantum state is implemented by projection measurements with single photon counters, which is traditionally explained by wavefunction collapse from Copenhagen interpretation. The quantum Darwinism experiment provides a further detailed mechanism to the emergent classical objectivity during the observation of quantum state (the photon 1), in the case of multiple observers (the photons 2 to 6). This mechanism is not only compatible with the Copenhagen interpretation of quantum measurement but also demonstrates a concrete Everett’s relative state \cite{everett1957rmp,everett1957rmp2} and thus consistent with the Many Worlds interpretation.}

In summary, we have experimentally observed the classical objectivity emerging from classical information redundancy of single quantum system on a six-qubit quantum Darwinism simulator. We have demonstrated that the environment acting as communication channel and selectively broadcasting quantum system's pointer states are the crucial mechanism of quantum Darwinism. Our work presents  {an essential} step to test the quantum Darwinism in small-scale controllable quantum environment. We expect further works to investigate the quantum Darwinism with more complex (e.g., larger-scale and mixed) quantum environment \cite{envNonideal2010pra,envNoisy2014prl} along with the considerable progress of current experimental quantum simulation technology \cite{simulatorPhoton,simulatorOnChip,simulatorTrapped,simulatorUltracold} and high-efficient quantum state characterization technology \cite{tomo2010,tomo2018}.

\textit{Acknowledgement}: This work was supported by the National Natural Science Foundation of China, the Chinese Academy of Sciences, the National Fundamental Research Program, the Anhui Initiative in Quantum Information Technologies and the Postdoctoral Innovation Talents Support Program.

\textit{Note}: After completing our work,  we became aware of two related experiments, one using two photons 
 {generating a cluster state \cite{pra2018} and the other one using single NV center \cite{unden2018revealing} to demonstrate quantum Darwinisim.}

\textit{Author Contributions}: M.-C.C., C.-Y.L., and J.-W.P. convinced and designed the work.M.-C.C., H.-S.Z., Y.L., D.W., X.-L.W., L.L., and N.-L.L. performed the experiment. M.-C.C., analyzed the data and wrote the manuscript with input from all authors. C.-Y.L. and J.-W.P. supervised the project.

\bibliographystyle{naturemag}
\bibliography{mybibtex}

\section*{Supplementary Information}

\textbf{Quantum states measurement and post-processing.}
The Darwinism states are measured through full quantum state tomography. There are $3^6$ measurement settings of Pauli operators ${\left\{ {{\sigma _X},{\sigma _Y},{\sigma _Z}} \right\}^{ \otimes 6}}$ and in each setting, we collect about 700 six-photon coincidence counts. Due to statistical fluctuation, the direct synthesis of  density matrix will generate some small negative eigenvalues and conflicts with the subsequent calculation of the entropy of information. We use the high-efficient algorithm introduced in \cite{fastMLE} to renormalize the eigenvalues. We further resample the measured density matrix and produce 100 new density matrices through Monte Carlo bootstrapping method. The 100 statistical-fluctuation density matrices are used to calculate all the presented error bars.

\end{document}